\documentclass[sigconf]{acmart}

\usepackage{booktabs} 
\usepackage[english]{babel}
\usepackage{subcaption}
\usepackage{mathrsfs}

\usepackage{multirow}
\usepackage{multicol}
\usepackage{array}
\usepackage{arydshln}

\usepackage{pgfplots}
\usepackage{siunitx}
\usepackage[inline]{enumitem}
\newlist{inlinelist}{enumerate*}{1}
\setlist*[inlinelist,1]{%
  label=(\roman*),
}
\newcommand\floor[1]{\lfloor#1\rfloor}

\newcommand\nrmf{NRM-F}

\newcommand\Hone{The ad-hoc retrieval performance of \nrmf~improves as we incorporate multiple document fields}

\newcommand\Htwo{\nrmf~performs better than competitive baselines, such as term matching and learning to rank}

\newcommand\Hthree{Learning a multiple-field document representation is superior to scoring based on individual field representations and summing}

\newcommand\Hfour{Learning per-field query representations performs better than learning a single query representation}

\newcommand\Hfive{The additional techniques of field-level masking and field-level dropout yield additional performance improvements}

\copyrightyear{2018}
\acmYear{2018}
\setcopyright{acmcopyright}
\acmConference{WSDM '18}{}{February 5--9, 2018, Los Angeles, CA.}

\author{Hamed Zamani}
\authornote{Work done while at Microsoft.}
\affiliation{%
  \institution{University of Massachusetts Amherst}
}
\email{zamani@cs.umass.edu}
\author{Bhaskar Mitra}
\authornote{The author is a part-time PhD student at UCL.}
\affiliation{%
  \institution{Microsoft and UCL}
}
\email{bmitra@microsoft.com}
\author{Xia Song}
\affiliation{%
  \institution{Microsoft}
}
\email{xiaso@microsoft.com}

\author{Nick Craswell}
\affiliation{%
  \institution{Microsoft}
}
\email{nickcr@microsoft.com}

\author{Saurabh Tiwary}
\affiliation{%
  \institution{Microsoft}
}
\email{satiwary@microsoft.com}

\begin{document}
\title{Neural Ranking Models with Multiple Document Fields}

\begin{abstract}

Deep neural networks have recently shown promise in the \emph{ad-hoc retrieval} task. However, such models have often been based on one field of the document, for example considering document title only or document body only. Since in practice documents typically have multiple fields, and given that non-neural ranking models such as BM25F have been developed to take advantage of document structure, this paper investigates how neural models can deal with multiple document fields. We introduce a model that can consume short text fields such as document title and long text fields such as document body. It can also handle multi-instance fields with variable number of instances, for example where each document has zero or more instances of incoming anchor text. Since fields vary in coverage and quality, we introduce a masking method to handle missing field instances, as well as a field-level dropout method to avoid relying too much on any one field. As in the studies of non-neural field weighting, we find it is better for the ranker to score the whole document jointly, rather than generate a per-field score and aggregate. We find that different document fields may match different aspects of the query and therefore benefit from comparing with separate representations of the query text. The combination of techniques introduced here leads to a neural ranker that can take advantage of full document structure, including multiple instance and missing instance data, of variable length. The techniques significantly enhance the performance of the ranker, and also outperform a learning to rank baseline with hand-crafted features.

\end{abstract}
\maketitle

\section{Introduction}
\label{sec:intro}
Deep neural networks have shown impressive performance in many machine learning tasks, including information retrieval models for ranking documents \cite{Dehghani:2017,Guo:2016,Huang:2013,Mitra:2017,Shen:2014,Xiong:2017}. These deep neural ranking models (NRMs) often consider a single source of document description, such as document title \cite{Huang:2013,Shen:2014} or body text \cite{Dehghani:2017,Mitra:2017}. However, in many retrieval scenarios, additional sources of document descriptions may be available. For instance in web search, each document consists of text fields specified by the document's HTML tags, such as title and body, as well as external sources of meta-information, such as the anchor text from incoming hyperlinks or the query text for which the document has been previously viewed. 

Learning a document representation suitable for retrieval tasks can be challenging when multiple document fields should be considered. These challenges primarily stem from the distinct properties of these diverse field types:
\begin{enumerate*}[label=(\roman*)]
  \item while the body of a web page is often long, the content of many other fields, such as title, are typically only a few terms in length,
  \item while some fields (e.g., body) contain a single instance of text, other fields may contain bags of multiple short texts (e.g., anchor text),
  \item multi-instance fields generally contain variable number of instances, e.g., zero or more instances of incoming anchor text for a given document,
  \item some fields, such as URL, may not contain natural language text, and finally
  \item fields vary in coverage and accuracy, for example a field that memorizes past queries that led to a click on the document may provide a very useful (high-accuracy) ranking signal \cite{Agichtein:2006}, but the coverage of that field may be relatively low because not every document has been clicked before
\end{enumerate*}. Each of these challenges increases the complexity of the representation learning task for documents with multiple fields. However, multiple fields associated with each document may contain complementary information that has motivated us to learn representation for documents by considering multiple fields in order to improve the retrieval performance.

\bigskip
In this paper, we propose \nrmf\footnote{The naming is inspired by BM25F \cite{Robertson:2004}.}, a \emph{general} framework for learning multiple-field document representation for ad-hoc retrieval. \nrmf~is designed to address the aforementioned challenges. More specifically, \nrmf~can handle multiple fields, both with single and multiple instances. In \nrmf, although the neural network parameters are shared among multiple instances of the same field, they are distinct across fields. This enables \nrmf~to uniquely model the content of each field based on its specific characteristics. 

We employ the same topology for the sub-networks corresponding to the different fields. However, there are a number of controlling hyper-parameters that determine the exact sub-network configuration for each field.
We introduce field-level masking to better cope with variable length inputs, i.e., fields with variable number of text instances. We also propose a novel field-level dropout technique that effectively regularizes the network and prevents it from over-dependence on high-accuracy fields, such as clicked queries. Given the intuition that different fields may match different aspects of the query, our model learns different query representations corresponding to different document fields.

We evaluate our models in the context of web search, using the queries sampled from the Bing's search logs. We study five fields in our experiments: title (single short text), body (single long text), URL (single short text, but not in a natural language), anchor texts (multiple short texts), and clicked queries (multiple short texts providing a ranking signal with relatively high accuracy). We consider this effective and diverse set of fields to make our findings more likely to generalize to other combinations of document fields.

In this work, we study the following research hypotheses:
\begin{description}
    \item[H1] \Hone.
    \item[H2] \Htwo.
    \item[H3] \Hthree.
    \item[H4] \Hfour.
    \item[H5] \Hfive.
\end{description}

Our experiments validate all these hypotheses, and investigate the effectiveness of our overall \nrmf~framework.

\vspace{-0.2cm}
\section{Related Work}
\label{sec:rel}

\subsection{Retrieval with Multiple Fields}

Information retrieval tasks may involve semi-structured data, meaning that the text of each document is divided into sections. Given a sufficiently fine-grained structure, some past research has studied the retrieval of the particular sections that best satisfy the user's query, such as in the INEX XML retrieval initiative \cite{Fuhr:2006,Lalmas:2007}. In web search it is more typical to consider coarse-grained sections such as title and body, also referred to as \emph{fields}, and use them to generate features in a document ranking task.

Using evidence from structure to improve document retrieval is well studied in information retrieval. \citet{Wilkinson:1994} proposed a number of hypotheses about how to combine section-level and document-level evidence. For example, taking the maximum section score, or a weighted sum of section scores, and then potentially combining with a document-level score. \citet{Robertson:2004} further proposed BM25F, an extension to the original BM25 model~\cite{Robertson:1994}, arguing that the linear combination of field-level scores is ``dangerous'', because it bypasses the careful balance across query terms in the BM25 model. The BM25F solution is to first combine frequency information across fields on a per-term basis, then compute a retrieval score using the balanced BM25 approach.

There are a number of alternative approaches to BM25F for the multiple-field document retrieval task. For instance, \citet{Piwowarski:2003} proposed a model based on Bayesian networks for retrieving semi-structured documents. \citet{Myaeng:1998} extended the InQuery retrieval system to semi-structured documents. \citet{Svore:2009} proposed a supervised approach, called LambdaBM25, that learns a BM25-like retrieval model based on the LambdaRank algorithm \cite{Burges:2007}. LambdaBM25 can also consider multiple document fields, without resorting to a linear combination of per-field scores. Dealing with multiple document fields without a linear combination was also studied by \citet{Ogilvie:2003}, who proposed and tested various combinations for a known-item search task, using a language modeling framework. \citet{Kim:2009} proposed a probabilistic model for the task of XML retrieval. Later on, \citet{Kim:2012} introduced a model based on relevance feedback for estimating the weight of each document field. 

\vspace{-2mm}
\subsection{Neural Networks for Ranking}
Several recent studies have applied deep neural network methods to various information retrieval applications, including question answering~\citep{Yang:2016}, click models~\citep{Borisov:2016}, ad-hoc retrieval \cite{Dehghani:2017,Mitra:2017,Xiong:2017}, and context-aware ranking~\citep{Zamani:2017b}. Neural ranking models can be partitioned into early and late combination models \cite{Dehghani:2017}. They can also be categorized based on whether they focus on lexical matching or learning text representations for semantic matching \cite{Mitra:2017}.

The early combination models are designed based on the interactions between query and document as the networks' input. For instance, the deep relevance matching model \cite{Guo:2016} gets histogram-based features as input, representing the interactions between query and document. DeepMatch \cite{Lu:2013} is another example that maps the input to a sequence of terms and computes the matching score using a feed-forward network. The local component of the duet model in \cite{Mitra:2017} and the neural ranking models proposed in \cite{Dehghani:2017,Xiong:2017} are the other examples for early combination models.

The late combination models, on the other hand, separately learn a representation for query and document and then compute the relevance score using a matching function applied on the learned representations. DSSM \cite{Huang:2013} is an example of late combination models that learns representations using feed-forward networks and then uses cosine similarity as the matching function. DSSM was further extended by making use of convolutional neural networks, called C-DSSM~\citep{Shen:2014}. The distributed component of the duet model \cite{Mitra:2017} also uses a similar architecture for learning document representation. We refer the reader to \cite{mitra2017introduction} that provides an overview of various (deep) neural ranking models.

In all of the aforementioned work, each document is assumed to be a single instance of text (i.e., single field).
However, documents often exist in a semi-structured format. In this paper, we focus on late combination models and propose a neural ranking model that takes multiple fields of document into account. Given the hypothesis provided in \cite{Mitra:2017}, our neural model can be further enriched by making use of lexical matching in addition to distributed matching. We leave the study of lexical matching for the future and focus on document representation learning.

\vspace{-2mm}
\section{The \nrmf~Framework}
\label{sec:ranking_model}
In this section, we first provide our motivation for studying the task of representation learning for documents with multiple fields, and formalize the task. We then introduce a high-level overview of our framework, and further describe how we implement each component of the proposed framework. We finally explain how we optimize our neural ranking model.

\vspace{-2mm}
\subsection{Motivation and Problem Statement}
\label{sec:method:motivation}

In many retrieval scenarios, there exist various sources of textual information (\emph{fields}) associated with each document $d$. In web search in particular, these sources of information can be partitioned into three categories. The first category includes the information provided by the structure and the content of document $d$ itself. Different elements of the web page specified by the HTML tags, e.g., title, header, keyword, and body, as well as the URL are examples of fields of this type. 
The second category includes the information provided by the other documents for representing $d$. For instance, when there is a hyperlink from document $d'$ to $d$, the corresponding anchor text may provide useful description of $d$. The third category contains information that we can infer from interactions between the retrieval system and its users. For instance in web search, when a user clicks on the document $d$ for a query $q$, the text of query $q$ can be used to describe $d$. \citet{Svore:2009} refer to these last two categories as popularity fields.

There are several previous studies showing that different fields may contain complementary information \cite{Robertson:2004,Svore:2009}. Therefore, incorporating multiple fields can lead to more accurate document representation and better retrieval performance. For example, clicked queries are highly effective for the retrieval tasks \cite{Agichtein:2006,Svore:2009,Xue:2004}. A number of prior studies \cite{Robertson:2004,Svore:2009}, have also investigated the usefulness of anchor texts for web search. However, for fresh or less popular documents that may not have enough anchor or clicked query text associated with them, the body text provides important description of the document. Similarly, the URL field may be useful for matching when the query expresses an explicit or implicit intent for a specific domain.
These complementary and diverse sources of textual descriptions have motivated us to study representation learning for ad-hoc retrieval by incorporating multiple fields.

The unique properties of these diverse document fields, however, make it challenging to model them within the same neural architecture. For example, the vocabulary and the language structure of clicked queries may be distinct from those of the body text, and in turn both may be distinct from the URL field. The document body text may contain thousands of terms, while the text in other fields may be only few terms in length. Finally, a key challenge also stems from the fact that a number of fields consist of multiple instances. For example, there are multiple anchor texts for each document $d$, and multiple queries can be found that previously led users to click on document $d$. A neural ranking model that considers these fields for document ranking must handle variable number of text instances per document field. To formulate the task, let $\mathcal{F}_d = \{F_1, F_2, \cdots, F_k\}$ denote a set of fields associated with the document $d$. Each field $F_i$ consists of a set of instances $\{f_{i1}, f_{i2}, \cdots, f_{im_i}\}$ where $m_i$ denotes the number of instances in the field $F_i$. The task is to learn a function $\Phi_D(\mathcal{F}_d)$ whose output is a representation for document $d$, suitable for the ad-hoc retrieval task.

\begin{figure}
    \centering
    \includegraphics[trim={11cm 6.4cm 11cm 4.5cm},clip,width=0.9\linewidth]{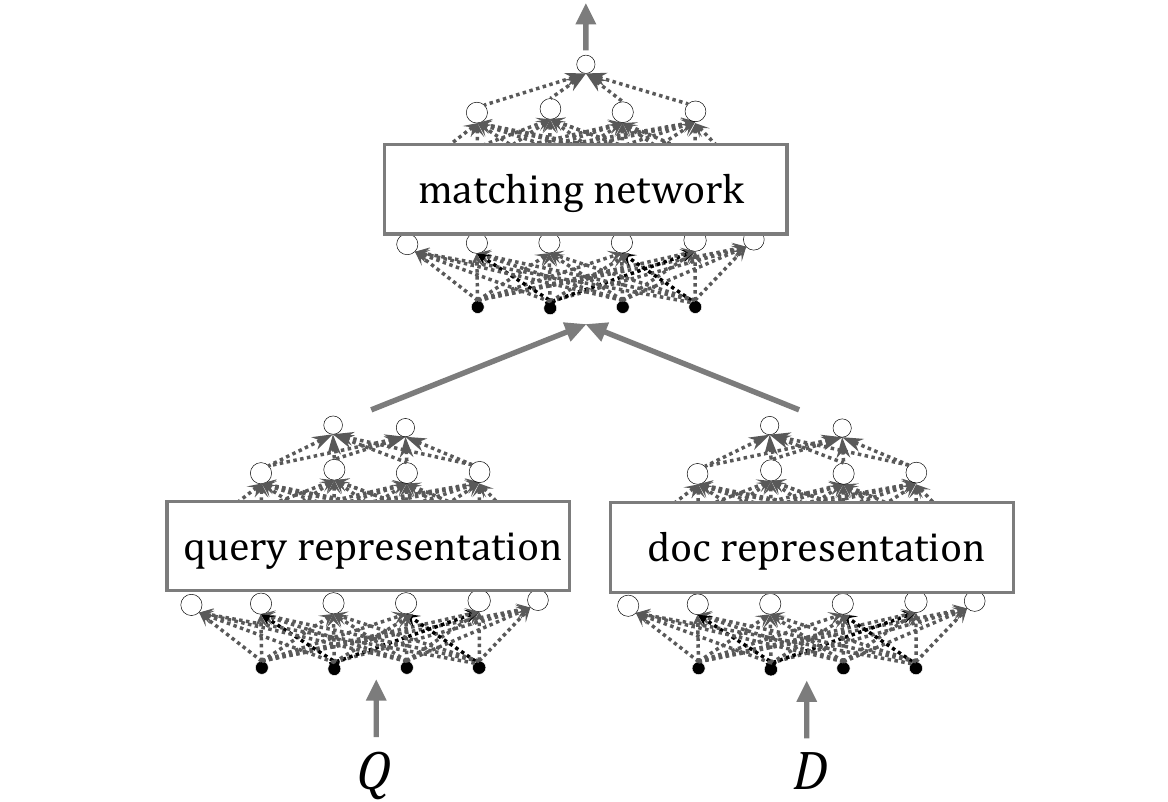}
    \caption{A neural ranking model architecture that consists of three major components: query representation, document representation, and matching network.}
    \label{fig:neural_ranking}
    \vspace{-4mm}
\end{figure}

\vspace{-2mm}
\subsection{High-Level Overview of the Framework}
\label{sec:method:overview}
In this paper, as shown in \figurename~\ref{fig:neural_ranking}, we focus on a late-combination and representation-focused neural ranking model. This architecture consists of three major components: document representation ($\Phi_D$), query representation ($\Phi_Q$), and the matching network ($\Psi$) which takes both representations and computes the retrieval score (i.e., $\text{score} = \Psi(\Phi_Q, \Phi_D)$).
In this section, we describe the high-level architecture used for the document representation network, which is the focus of the paper. Sections~\ref{sec:method:query} and \ref{sec:method:matching} review how the query representation and matching network components are respectively implemented.

\begin{figure*}
    \centering\vspace{-8mm}
    \includegraphics[trim={0 3cm 0 3.5cm},clip,width=0.9\textwidth]{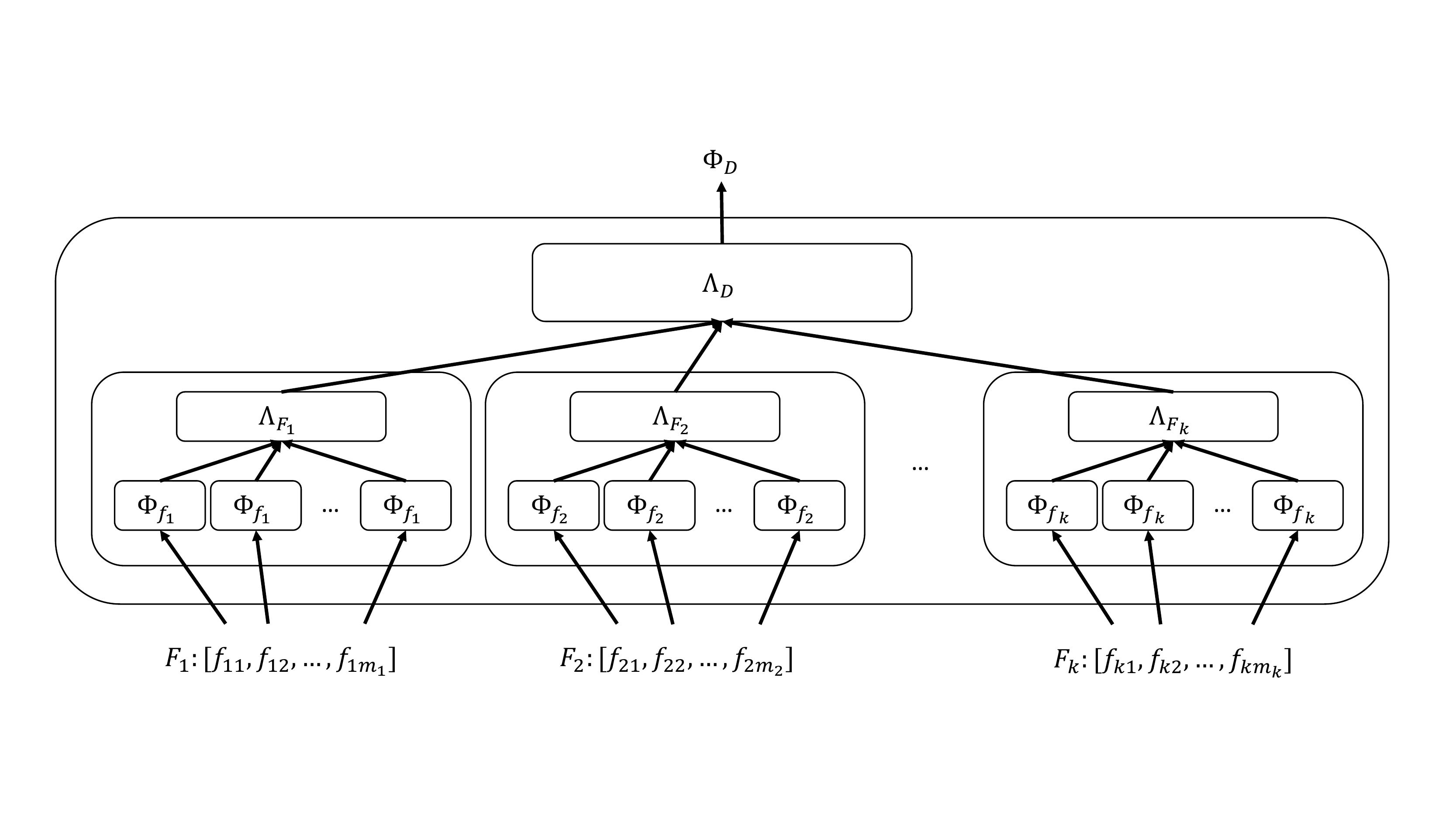}
    \vspace{-4mm}
    \caption{The high-level architecture of our document representation framework. In this architecture, aggregating field-level representations using $\Lambda_D$ produces the document representation $\Phi_D$. The representation for the $i^{th}$ field is computed by aggregating ($\Lambda_{F_i}$) the representations learned for the instances of the field using $\Phi_{f_i}$.}
    \label{fig:framework}
    \vspace{-4mm}
\end{figure*}

To learn multiple-field document representation (i.e., $\Phi_D$), the framework first learns a representation for each individual instance in a field. The framework then aggregates these learned vector representations to represent the field as a whole. It finally aggregates all the field specific representations for the document. This framework is visualized in \figurename~\ref{fig:framework}. 

To formally describe our framework, the document representation learning function $\Phi_D$ can be calculated as:
\begin{equation}
    \Phi_D(\mathcal{F}_d) = \Lambda_D(\Phi_{F_1}(F_1), \Phi_{F_2}(F_2), \cdots, \Phi_{F_k}(F_k))
\end{equation}
where $\Phi_{F_i}$ denotes the representation learning function for the field $F_i$. Note that the representation learning functions differ for different fields, since the fields have their own unique characteristics and need their own specific functions. $\Lambda_D$ aggregates representations learned for all the fields. Each $\Phi_{F_i}$ is also calculated as:
\begin{equation}
    \Phi_{F_i}(F_i) = \Lambda_{F_i}(\Phi_{f_i}(f_{i1}), \Phi_{f_i}(f_{i2}), \cdots, \Phi_{f_i}(f_{i m_i}))
\end{equation}
where $\Phi_{f_i}$ denotes the representation learning function for each instance of the $i^{th}$ field (e.g., each anchor text). Note that $\Phi_{f_i}$ is the same function for all the instances of a given field. The function $\Lambda_{F_i}$ aggregates the representation of all instances in the $i^{th}$ field. 

To summarize, our document representation framework consists of three major components: learning representation for an instance of each field (i.e., $\Phi_{f_i}$), field-level aggregation (i.e., $\Lambda_{F_i}$), and document-level aggregation (i.e., $\Lambda_D$). Sections~\ref{sec:method:instrance} and \ref{sec:method:aggregation} describe how we define or learn these functions.

\begin{figure}
    \centering
    \includegraphics[trim={7.5cm 3cm 6.5cm 3cm},clip,width=\linewidth]{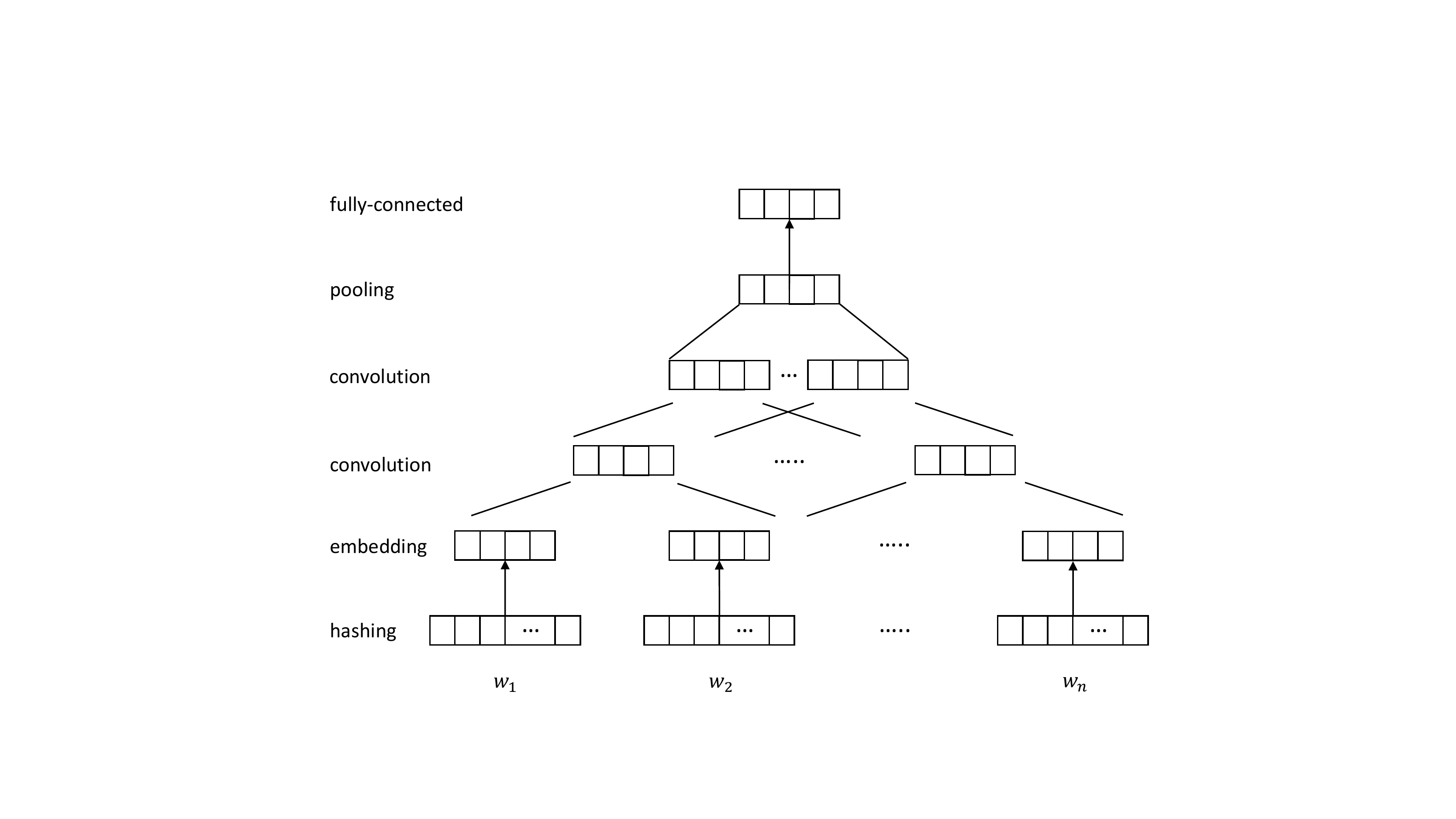}
    \vspace{-6mm}
    \caption{Instance-level representation learning network. This model embeds the character n-gram representation of each word $w_i$ which is followed by two 1-D convolutional layers. The outputs of the second set of convolutional operations are pooled and then fed to a fully-connected layer to compute the final representation for each instance of a field.}
    \label{fig:network}
    \vspace{-2mm}
\end{figure}

\vspace{-2mm}
\subsection{Instance-Level Representation Learning}
\label{sec:method:instrance}

In this subsection, we describe our neural architecture for learning representations of individual text instances in a document field. In particular, we explain how the functions $\Phi_{f_i}$ are implemented. As pointed out in Section~\ref{sec:method:motivation}, each field has its own unique characteristics. One approach would be to use different neural architectures for different fields. However, in the interest of proposing a general framework, we choose an architecture that can be used for all fields, but the exact configurations are controlled by a set of hyper-parameters specified per field. These hyper-parameters are selected for each field individually based on a validation set.

\figurename~\ref{fig:network} shows the design of the per-instance model architecture. In our architecture, each term is represented using a character n-gram hashing vector, introduced by Huang et al.~\cite{Huang:2013}. These are extremely sparse vectors whose dimensions correspond to all possible character n-grams. Therefore, we represent the input layer of our network using sparse tensors which is memory-efficient and also improves the efficiency of the model. Similar to \cite{Huang:2013,Shen:2014}, $n$ was set to $3$ in our experiments which causes limited number of term hash collisions. We use the character n-gram representation for the following reasons: (1) it can represent out of vocabulary terms, and (2) the number of all possible tri-grams is much lower than the term-level vocabulary size which significantly reduces the number of parameters needed to be learned by the network. We use a linear embedding layer to map a character n-gram representation to a dense low-dimensional representation by multiplying the sparse input tensor for each term and an embedding matrix $\mathcal{E} \in \mathbb{R}^{N \times l}$ where $N$ is total number of possible n-grams and $l$ denotes the embedding dimensionality. The output of this layer for each word is normalized to prevent over-weighting long words, and represents relevance-based word embedding \cite{Zamani:2017a}. Inspired by C-DSSM \cite{Shen:2014} and the Duet Model \cite{Mitra:2017}, this layer is followed by a one-dimensional convolution layer. The aim of this layer is to capture the dependency between terms. 
We further use an additional convolution layer whose window size is set to be larger for the body field to capture sentence-level representations, and smaller for the short texts fields. We pool the output of the second convolution layer which is followed by a fully-connected layer to compute the final representation for an instance of a given field. The choice of max-pooling and average-pooling is a hyper-parameter in our model. In this network, we use dropout \cite{Srivastava:2014} to avoid over-fitting.

\vspace{-2mm}
\subsection{Aggregating Representations}
\label{sec:method:aggregation}
As shown in \figurename~\ref{fig:framework}, \nrmf~consists of two sets of aggregation components: $\Lambda_{F_i}$ and $\Lambda_{D}$. $\Lambda_{F_i}$ aggregates the representations learned for the instances of a specific field. For each multi-instance field $F_i$, we select a bag of at most $M_i$ instances, and use zero padding when less than $M_i$ instances are available. $\Lambda_{F_i}$ averages the individual representations learned for the instances of the field~$F_i$.

The component $\Lambda_{D}$ aims at aggregating the representations learned for different fields. To be able to learn different query representations for each field, $\Lambda_{D}$ only concatenates the input vectors to be served in the matching function explained in Section~\ref{sec:method:matching}.

\vspace{-2mm}
\subsection{Field-Level Masking}
\label{sec:method:masking}
The number of instances in multi-instance fields, such as anchor text, varies across documents. As shown in \tablename~\ref{tab:data}, a significant number of documents may not contain any anchor text or clicked queries. To deal with such cases, as mentioned in Section~\ref{sec:method:aggregation}, we use zero padding. Although padding is a popular approach and has been previously used in neural ranking models \cite{Dehghani:2017,Mitra:2017,Shen:2014}, it suffers from a major drawback: by doing padding, the network assumes that a part of the input vector is zero; however padding represents missing values. In the extreme case, assume that there is no available anchor text for a given document; therefore, the input for the anchor text field is all zero. The gradients, however, are not zero (because of the bias parameters). This means that the back-propagation algorithm updates the weights for the sub-network corresponding to the anchor text field; which is not desirable---we do not want to update the weights when the input data is missing. This becomes crucial when there are many missing values in training data, similar to our task.

To tackle this problem, we propose a simple approach, called \textit{field-level masking}. Let $R_i \in \mathbb{R}^{M_i \times D_i}$ denote the representation learned for the $i^{th}$ field (i.e., the output of $\Lambda_{F_i}$) where $M_i$ and $D_i$ respectively represent the maximum number of instances (fixed value) and the dimensionality for instance representation. We generate a binary masking matrix $B_i \in \mathbb{B}^{M_i \times D_i}$ whose rows are all zero or all one, showing whether each field instance exists or is missing. In masking, we use $R_i \circ B_i$ (i.e., element-wise multiplication) as the representation for $F_i$. 
We multiply the representations for existing field instances by one (means no change) and those for the missing instances by zero. This not only results in zero representation for missing values, but also forces the gradients to become zero. Therefore, the back-propagation algorithm does not update the weights for the sub-networks corresponding to missing values.

The masking matrix is also useful for computing the average in $\Lambda_{F_i}$ (see Section~\ref{sec:method:aggregation}). Averaging is a common approach for aggregating different representations, such as average word embedding for query representation \cite{Zamani:2016} and neural ranking models \cite{Dehghani:2017}. However, in case of variable length inputs, averaging penalizes short inputs which are padded by zero. To address this issue, we can compute the exact average vector by summing the inputs and dividing them by the summation over the masking matrix. Note that the masking technique should be applied at both training and testing times.

\begin{table*}[t]
    \centering
    \caption{Statistics and characteristics of the document fields used in our experiments.}\vspace{-4mm}
    \begin{tabular}{llll}\hline\hline
        \textbf{Fields} & \textbf{Type} & \textbf{Coverage} & \textbf{Specific Feature} \\\hline
        Title & Single Instance & $100\%$  & Short text. \\
        URL & Single Instance & $100\%$ & Short text, but not in a natural language. \\
        Body & Single Instance & $100\%$ & Long text. \\
        Anchor texts & Multiple Instance & $61\%$ & Short texts with relatively low coverage. \\
        Clicked queries & Multiple Instance & $73\%$ & Short texts with relatively low coverage. A high-accuracy field. \\\hline\hline
    \end{tabular}
    \label{tab:data}
\end{table*}

\vspace{-2mm}
\subsection{Field-Level Dropout}
\label{sec:method:dropout}
As widely known and also demonstrated in our experiments, clicked queries is an effective field for representing documents in the retrieval task \cite{Agichtein:2006,Svore:2009}. When such a high-accuracy field is available, there is a risk that the network relies on that field, and pays less attention to learning proper representations for the other fields. This can lead to poor performance of the model when the high-accuracy field is absent (low coverage).

Although we use dropout in our neural ranking model (see Section~\ref{sec:method:instrance}), it is not sufficient for the task of document representation learning with multiple document fields, in particular when at least a dominant input field exists.
To regularize the network in such cases, we propose a simple \textit{field-level dropout} technique---randomly dropping all the units corresponding to a field. In other words, we may randomly drop, say, the clicked queries field or the body field at training time to prevent the neural ranking model from over-dependence on any single field. This approach is back-propagation friendly (all the proofs presented in \cite{Srivastava:2014} are applicable to the field-level dropout). Field-level dropout contains $k$ hyper-parameters, where $k$ denotes the total number of fields and each parameter controls the probability of keeping the corresponding field. Note that dropout only happens at the training time and all the units are kept at the validation and test times.

\vspace{-2mm}
\subsection{Query Representation}
\label{sec:method:query}
Since in this paper we focus on the ad-hoc retrieval task, the only available information for the query is the query text. Therefore, to represent the query (i.e., $\Phi_Q$), we use the same network architecture as the one used for each instance of a document field (see Section~\ref{sec:method:instrance}). Note that different document fields may match with different aspects of a query. Therefore, the output dimensionality of the query representation network is equal to the sum of the dimensions for all fields' representations. In other words, \nrmf~learns different representations of the query for each document field. 

\vspace{-2mm}
\subsection{Matching Network}
\label{sec:method:matching}
In this subsection, we describe how we compute the retrieval score given the output of query representation and document representation networks (i.e., the function $\Psi$). To do so, we compute the Hadamard product of the representations; which is the element-wise product of two matrices with the same dimensionality. We then use a fully-connected neural network with a single non-linear hidden layer to compute the final retrieval score. We avoid computing dot product or cosine similarity which would reduce the contribution of each field to a single score, forcing us to combine them linearly which is less effective as demonstrated by \citet{Robertson:2004} and our results in Section~\ref{sec:exp:results}.

\vspace{-2mm}
\subsection{Training}
\label{sec:method:training}
We use a pairwise setting to train the designed neural ranking model. Let $T = \{(q_1, d_{11}, d_{12}, y_{11}, y_{12}), (q_2, d_{21}, d_{22}, y_{21}, y_{22}), \cdots, (q_n, d_{n1},\\d_{n2}, y_{n1}, y_{n2})\}$ be a set of $n$ training instances. Each training instance consists of a query $q_i$, two documents $d_{i1}$ and $d_{i2}$, as well as their corresponding labels $y_{i1}$ and $y_{i2}$. 
We consider cross entropy loss function to train neural ranking models:
\begin{equation*}
    \mathcal{L} = -\frac{1}{|T|} \sum_{i=1}^{|T|}{\frac{g(y_{i1})}{g(y_{i1})+g(y_{i2})} \log p_{i1} + \frac{g(y_{i2})}{g(y_{i1})+g(y_{i2})} \log (1-p_{i1})}
\end{equation*}
where $g(\cdot)$ is a gain function. We use an exponential gain function same as the one used in calculating NDCG. $p_{i1}$ is the estimated probability for $d_{i1}$ being more relevant than $d_{i2}$. $p_{i1}$ is calculated via softmax on the predicted labels: $p_{i1} = {\exp{(\hat{y}_{i1})}}/({\exp{(\hat{y}_{i1})} + \exp{(\hat{y}_{i2})}})$, where $\hat{y}_{i1}$ and $\hat{y}_{i2}$ denote the estimated scores for $d_{i1}$ and $d_{i2}$, respectively.


\vspace{-2mm}
\section{Experiments}
\label{sec:exp}


\subsection{Data}
\label{sec:exp:data}
To evaluate our models, we randomly sampled ${\sim}140k$ queries from the Bing's search logs for the English United States market from a one-year period. For each query, the documents returned by the Bing's production ranker in addition to those retrieved by a diverse set of experiments were labelled by human judges on a five-point scale: perfect, excellent, good, fair, and bad. In total, the data consists of ${\sim}3.8$ million query-document pairs which was randomly partitioned into three sets---$80\%$ for training, $10\%$ for validation, and $10\%$ for testing---such that no distinct query appears in more than one set. Similar to \cite{Huang:2013,Mitra:2017,nalisnick2016improving}, we evaluate all models under the \emph{telescoping} setting by re-ranking the candidate documents for each query. Since our neural ranking model is a pairwise learning to rank model, for each query we generate all possible $<q, d_1, d_2>$ triples such that the relevance label for $d_1$ and $d_2$ are different with respect to $q$. To avoid biasing towards the queries with many documents, at most $50$ triples per query were sampled for training based on a uniform distribution over all possible label pairs.

The contents of web pages were retrieved from the Bing's web index and were parsed using a proprietary HTML parser. We made sure that all the documents in our data contain title and body. All texts were normalized by lower-casing and removing non-alphanumerical characters. The URLs were split using a simple proprietary approach. We set the maximum length of $20$, $10$, $1000$, $10$, and $10$ for title, URL, body, anchor text, and clicked query, respectively. We used at most $5$ anchor texts and at most $5$ clicked queries per document.\footnote{The maximum number of instances per field can be set to a much larger value. Since the network parameters for instances of each field are shared and the inputs are represented as sparse tensors, increasing the maximum number of instances would have a minor memory effect.} They were selected based on a simple count-based functions; means that the most common anchor texts and clicked queries for each document were selected. The statistics of our data for each field is reported in \tablename~\ref{tab:data}.

\vspace{-2mm}
\subsection{Experimental Setup}
\label{sec:exp:setup}
All the models were implemented using TensorFlow\footnote{\url{http://tensorflow.org/}}. We used Adam optimizer \cite{Kingma:2014} to train our models. The learning rate was selected from $[1e-3, 5e-4, 1e-4, 5e-5, 1e-5]$. We set the batch size to $64$ and tuned the hyper-parameters based on the loss values obtained on the validation set. We selected the layer sizes from $\{100, 300, 500\}$ and the convolution window sizes from $\{1, 3, 10, 20,\\50\}$ for long texts (i.e., body) and from $\{1, 3, 5, 10\}$ for short texts (i.e., the other fields). The convolution strides were selected from $\{1, \floor{ws/4}, \floor{ws/2}, ws\}$ where $ws$ denotes the convolution window size. The keep probability parameters for both conventional and field-level dropouts were selected from $\{0.5, 0.8, 1.0\}$.

As explained in Section \ref{sec:method:instrance}, the input layer of the networks uses tri-gram hashing with ${\sim}50k$ dimensions, i.e, all possible character tri-grams with alphanumerical characters plus a dummy character for the start and the end of each word. The tri-gram embedding dimensionality (i.e., the first layer) was set to $300$. This embedding matrix is shared among all fields. Following \cite{Huang:2013,Mitra:2017,Shen:2014}, we used $\tanh$ as the activation function for all hidden layers.

We use NDCG at two different ranking levels (NDCG@1 and NDCG@10) to evaluate the models. The significance differences between models are determined using the paired t-test at a $95\%$ confidence level ($p\_value < 0.05$).

\vspace{-2mm}
\subsection{Experimental Results}
\label{sec:exp:results}
In this subsection, we empirically address the hypotheses mentioned in Section~\ref{sec:intro}.

\begin{table}[t]
    \centering
    \caption{Performance of the proposed framework with different fields. The superscript + shows significant improvements for the models with two fields compared to the ones with each of the fields, individually. The superscript * denotes significant improvements over all the other models.}\vspace{-2mm}
    \begin{tabular}{p{4cm}ll}\hline\hline
        \textbf{Field(s)} & \textbf{NDCG@1} & \textbf{NDCG@10} \\\hline
        Title & 0.4226 & 0.5883 \\
        URL & 0.4366 & 0.5865 \\
        Body & 0.4115 & 0.5850 \\
        Anchor texts & 0.4386 & 0.5933 \\
        Clicked queries & 0.4661 & 0.6116 \\\hline
        Title + URL & 0.4425\textsuperscript{+} & 0.6065\textsuperscript{+} \\
        Title + Body & 0.4316\textsuperscript{+} & 0.6098\textsuperscript{+} \\
        Title + Anchor texts & 0.4507\textsuperscript{+} & 0.6062\textsuperscript{+} \\
        Title + Clicked queries & 0.4680 & 0.6180\textsuperscript{+} \\\hline
        All & \textbf{0.4906}\textsuperscript{*} & \textbf{0.6380}\textsuperscript{*} \\\hline\hline
    \end{tabular}
    \label{tab:fields}
    \vspace{-2mm}
\end{table}

\begin{table*}[t]
    \centering
    \caption{Comparison of the proposed model with baselines for a single field (Title or Body). The superscripts denote significant improvements over the models specified by the ID column.}\vspace{-2mm}
    \begin{tabular}{llllll}\hline\hline
        \multirow{2}{*}{\textbf{ID}} & \multirow{2}{*}{\textbf{Model}} & \multicolumn{2}{c}{\textbf{Title}} & \multicolumn{2}{c}{\textbf{Body}} \\
        & & \textbf{NDCG@1} & \textbf{NDCG@10} & \textbf{NDCG@1} & \textbf{NDCG@10} \\\hline
        1 & BM25 & 0.4039 & 0.5752 & 0.3957 & 0.5693 \\
        2 & LTR & 0.4122 & 0.5861 & 0.3996 & 0.5792 \\
        3 & DSSM & 0.4112 & 0.5858 & 0.3961 & 0.5713 \\
        4 & C-DSSM & 0.4148 & 0.5874 & 0.3957 & 0.5695 \\
        5 & Duet (distributed) & 0.4164 & 0.5877 & 0.4066 & 0.5788 \\\hline
        6 & \nrmf~- Single Field & \textbf{0.4226}\textsuperscript{12345} & \textbf{0.5883}\textsuperscript{123} & \textbf{0.4115}\textsuperscript{12345} & \textbf{0.5850}\textsuperscript{12345} \\\hline\hline
    \end{tabular}
    \label{tab:single_baseline}
\end{table*}

\paragraph{\textbf{H1: \Hone.}}
In this set of experiments, we address our first hypothesis (H1) by evaluating our model with each single field individually, with field pairs with title, and finally with all the fields together. The results are reported in \tablename~\ref{tab:fields}. Although title, URL, and body have much higher coverage compared to anchor texts and clicked queries (see \tablename~\ref{tab:data}), the performances achieved by anchor texts and clicked queries are superior to the other fields.\footnote{We randomly shuffled the documents with equal retrieval scores for a query. This process was repeated for $10$ times and the average performance is reported.} Incorporating clicked queries demonstrates the highest performance. Pairing Title with any of the other field ``X'' leads to a better performance compared to Title and ``X'', individually. These improvements are statistically significant, except for NDCG@1 in Title+Clicked queries. The reason is that clicked queries are very effective for web search, especially for the first retrieved document. Adding title to clicked queries, however, significantly improves the search quality for the top $10$ documents. The \nrmf~model with all fields achieves the highest performance with statistically significant margins. This suggests that the proposed framework is able to learn a more accurate document representation for the ad-hoc retrieval task by considering multiple document fields; thus the hypothesis H1 is validated.

\paragraph{\textbf{H2: \Htwo.}}
To demonstrate that the proposed instance-level representation model performs reasonably well for both short and long texts, we first evaluate our models against a set of baselines using a single field, title only and body only. We consider the following baselines: BM25 \cite{Robertson:1994}, a state-of-the-art learning to rank model with hand-crafted features (LTR), DSSM \cite{Huang:2013}, C-DSSM \cite{Shen:2014}, and the distributed part\footnote{To have a fair comparison, we only consider the distributed part of the model. Note that all the listed neural models, including \nrmf, can be further enriched by using lexical matching, similar to the local part of the duet model.} of the duet model proposed by Mitra et al. \cite{Mitra:2017}.
The LTR baseline uses an internal advanced implementation of the LambdaMART algorithm \cite{Burges:2010} that has been used in the production. We used the features that have been typically extracted from query and document texts. Indeed, from those listed in \cite{Qin:2010}, we used all the features that can be extracted from query and title/body. 

To have a fair comparison, we trained all the models using the same training data and pairwise setting.\footnote{The original DSSM and C-DSSM models use binary labels (click data) and random negative sampling for training; however, as suggested by Mitra et al. \cite{Mitra:2017} using explicit judgments leads to a better performance compared to random negative sampling} The hyper-parameters in all the models, including the baselines, were optimized for Title and Body, separately. Due to the memory constraints, the C-DSSM and Duet cannot use ${\sim}50k$ tri-grams for the word hashing phase (only for Body). Therefore, as suggested in \cite{Mitra:2017}, we use top $2k$ popular n-grams for these models. Note that since our model use sparse tensors for word hashing, it is memory-efficient and does not have the same issue.\footnote{C-DSSM and Duet perform convolution on top of word hashing layer; thus, the word hashing phase cannot be implemented using sparse tensors (at least not supported by deep learning libraries, such as TensorFlow and CNTK).} 

The results for Title as an example of short text and Body as an example of long text are reported in \tablename~\ref{tab:single_baseline}. According to this table, the proposed method outperforms all the baselines for both Title and Body. The improvements are statistically significant in nearly all cases. This demonstrates the potential of our model to be used for both short and long texts. The improvements are higher for Body, which makes our model even more suitable for long text. This experiment suggests that our instance-level representation model performs reasonably well. 


\begin{table}[t]
    \centering
    \tabcolsep=0.09cm
    \caption{Performance of the proposed framework with all fields compared to baselines. The superscript * denotes significant improvements over all the other models.}\vspace{-2mm}
    \begin{tabular}{lll}\hline\hline
        \textbf{Model} & \textbf{NDCG@1} & \textbf{NDCG@10} \\\hline
        BM25-Field Concatenation & 0.4281 & 0.5953 \\
        BM25F & 0.4431 & 0.6020 \\
        LTR & 0.4888 & 0.6341 \\
        NRM-Field Concatenation & 0.4582 & 0.6110 \\\hdashline
        NRM-Score Aggregation-Ind. Training & 0.4729 & 0.6229 \\
        NRM-Score Aggregation-Co-training & 0.4743 & 0.6279 \\\hdashline
        \nrmf~-Single Query Representation & 0.4846 & 0.6345 \\\hline
        \nrmf & \textbf{0.4906}\textsuperscript{*} & \textbf{0.6380}\textsuperscript{*} \\\hline\hline
    \end{tabular}
    \label{tab:baseline}
    \vspace{-2mm}
\end{table}

To evaluate our model with multiple instances, we consider the following baselines: (1) BM25 by concatenating all the fields, (2) BM25F \cite{Robertson:2004} which has been widely used for ad-hoc retrieval with multiple document fields, (3) a learning to rank (LTR) model with hand-crafted features extracted from all the fields, and (4) our neural ranking model with concatenation of all fields as a single input text (i.e., NRM - Field Concatenation). Similar to the last experiments, for LTR we consider all the typical features that can be extracted from text inputs (among those listed in \cite{Qin:2010} for the LETOR dataset). The features were extracted for all the fields. The learning algorithm for LTR is the same as the one used in the previous experiment. All models were trained on the same training set, and their hyper-parameters were tuned on the same validation set. As shown in \tablename~\ref{tab:baseline}, \nrmf~significantly outperforms all the baselines. This suggests that \nrmf~not only eliminates the hand-crafted feature engineering for ad-hoc retrieval, but also learns an accurate document representation that leads to higher retrieval performance. The results also validate our second hypothesis.

\begin{table*}[t]
    \centering
    \caption{Investigating the effectiveness of field-level masking and dropout. The superscripts denote significant improvements over the models specified by the ID column.}\vspace{-4mm}
    \begin{tabular}{llp{2cm}p{2cm}p{2cm}p{2cm}}\hline\hline
        \multirow{2}{*}{\textbf{ID}} & \multirow{2}{*}{\textbf{Model}} & \multicolumn{2}{c}{\textbf{All fields}} & \multicolumn{2}{c}{\textbf{All fields except clicked queries}} \\
         & & \multicolumn{1}{c}{\textbf{NDCG@1}} & \multicolumn{1}{c}{\textbf{NDCG@10}} & \multicolumn{1}{c}{\textbf{NDCG@1}} & \multicolumn{1}{c}{\textbf{NDCG@10}} \\\hline
        1 & \nrmf~ (no masking, no dropout) & ~~~~~~~0.4818 & ~~~~~~~~0.6327 & ~~~~~~~0.4577 & ~~~~~~~~0.6152 \\
        2 & \nrmf~ with masking & ~~~~~~~0.4856\textsuperscript{1} & ~~~~~~~~0.6353\textsuperscript{1} & ~~~~~~~0.4602\textsuperscript{1} & ~~~~~~~~0.6174\textsuperscript{1} \\
        3 & \nrmf~ with masking \& dropout & ~~~~~~~\textbf{0.4906}\textsuperscript{12} & ~~~~~~~~\textbf{0.6380}\textsuperscript{12} & ~~~~~~~\textbf{0.4613}\textsuperscript{1} & ~~~~~~~~\textbf{0.6181}\textsuperscript{1} \\\hline\hline
    \end{tabular}
    \label{tab:dropout}
\end{table*}

\begin{table*}[t]
    \centering
    \caption{Performance analysis based on query length, dividing the test queries into three evenly-sized groups.}\vspace{-3mm}
    \begin{tabular}{p{2cm}llllll}\hline\hline
        \multirow{2}{*}{\textbf{Model}} & \multicolumn{2}{c}{\textbf{Short queries}} & \multicolumn{2}{c}{\textbf{Medium-length queries}} & \multicolumn{2}{c}{\textbf{Long queries}} \\
         & \textbf{NDCG@1} & \textbf{NDCG@10} & \textbf{NDCG@1} & \textbf{NDCG@10} & \textbf{NDCG@1} & \textbf{NDCG@10} \\\hline
        LTR & 0.5040 & 0.6470 & 0.4753 & 0.6332 & \textbf{0.4799} & 0.6162 \\
        \nrmf & \textbf{0.5132} & \textbf{0.6584} & \textbf{0.4846} & \textbf{0.6355} & 0.4723 & \textbf{0.6186} \\\hline\hline
    \end{tabular}
    \label{tab:qlen}
    \vspace{-3mm}
\end{table*}

\paragraph{\textbf{H3: \Hthree.}}
A simple approach for coping with multiple document fields is to calculate the matching score for the query and each of the document fields and then aggregate the scores. We tried two score aggregation methods, one learns a neural ranking model for each document field individually and then linearly interpolates their scores. Although the other one also interpolates the scores obtained by different fields, the neural networks for different fields are co-trained together. The results in \tablename~\ref{tab:baseline} show that co-training leads to a better performance compared to isolated training of the model for different fields, which is expected. The results also suggest that \nrmf~performs better than neural ranking models with score aggregation. The improvements are statistically significant. Therefore, this experiment validates our third hypothesis.

\paragraph{\textbf{H4: \Hfour.}}
As mentioned in Section~\ref{sec:method:query}, we believe that different aspects of the query can match different fields, and thus different query representations are needed for different fields. Our empirical results in \tablename~\ref{tab:baseline} also validate this hypothesis by showing that \nrmf~provides a superior performance in comparison with exactly the same neural ranking model, but with single query representation for different fields.

\paragraph{\textbf{H5: \Hfive.}}
To study this hypothesis, we report the results for the following models: (1) our neural ranking model with no field-level masking and dropout, (2) our model with only field-level masking, and eventually (3) our model with both field-level masking and dropout. Note that all the models use conventional dropout~\cite{Kingma:2014}. \tablename~\ref{tab:dropout} reports the results for all fields and for all fields except clicked queries. According to this table, field-level masking is useful to cope with multi-instance fields and significantly improves the performance. The model with both field-level masking and dropout achieves the highest performance; however, the field-level dropout technique is significantly helpful, when at least one of the fields is dominant (i.e., the high-accuracy fields like clicked queries).

\begin{figure}
    \centering
    \vspace{-6mm}
    \includegraphics[width=0.7\linewidth]{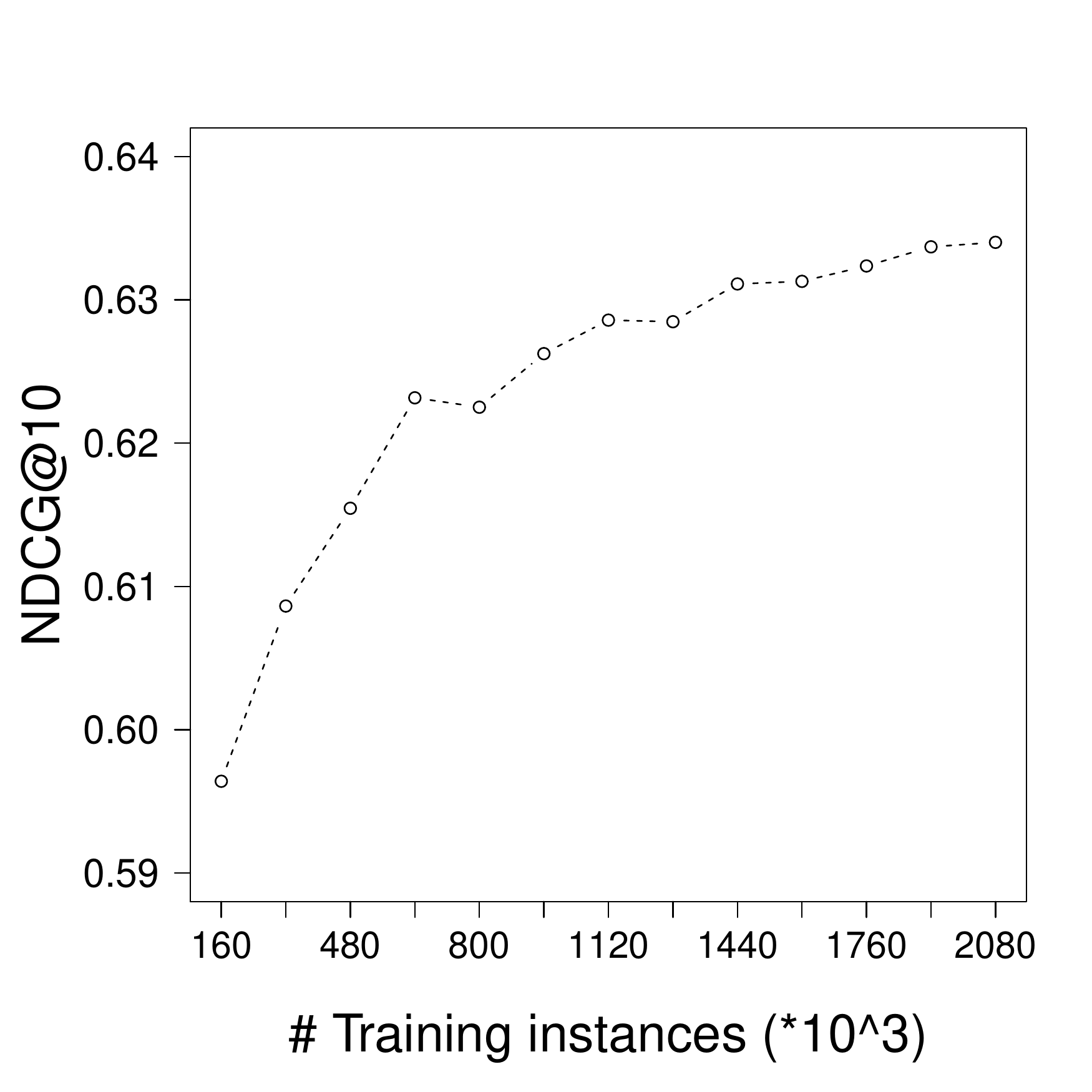}
    \vspace{-3mm}
    \caption{Learning curve demonstrating the performance of NRM-F in terms of NDCG@10 with respect to the size of training set.}\vspace{-2mm}
    \label{fig:learning_curve}
\end{figure}

\vspace{-2mm}
\subsection{Additional Analysis}

\paragraph{\textbf{Learning curve.}}
It has always been important to know how much data is needed to train the model. We plot the learning curve for our \nrmf~model with all fields in \figurename~\ref{fig:learning_curve}. The performance is reported in terms of NDCG@10 on the test set. According to this figure, we need approximately two million training instances to have a relatively stable performance.


\paragraph{\textbf{Analysis by query length.}}
In this analysis, we uniformly split the test queries into three buckets based on their query length. Therefore, the number of queries in the buckets are approximately equal. The first bucket includes the shortest and the last one includes the longest queries. The results for \nrmf~and the LTR baseline with all fields (the one used in \tablename~\ref{tab:baseline}) are reported in \tablename~\ref{tab:qlen}. According to this table, our improvements over the LTR baseline generally decrease by increasing the query length. In other words, \nrmf~performs relatively better for shorter queries. The reason is that long queries are often rare and thus it is likely that the models based on representation learning work much better for shorter queries. On the other hand, 
the experiment is in a telescoping setting with anchor texts and clicked queries. Therefore, the additional terms and synonyms provided by, let say, clicked queries empower the LTR method that uses term matching features. In addition, for long queries in a telescoping setting, ignoring a query term is relatively likely to do not harm the results.

\vspace{-2mm}
\section{Conclusions and Future Work}
\label{sec:conclusion}
In this paper, we proposed \nrmf, a \emph{general} framework for the task of multiple-field document representation learning for ad-hoc retrieval. \nrmf~can consume both short text and long text fields. It can also handle the multi-instance fields, such as anchor text. Since fields vary in coverage, we introduced a field-level masking method to handle missing field instances. We also proposed a field-level dropout technique to prevent the model from over-dependence on high-accuracy fields, such as clicked queries. We performed extensive experiments using a large set of query-document pairs labelled by human judges. Our experiments suggested that incorporating multiple document fields significantly improves the performance of neural ranking models by a large margin. Our model also outperforms state-of-the-art traditional term matching and learning to rank models, significantly. We showed that multiple query representations are needed for different document fields, based on the intuition that different aspects of a query may be matched against different fields of a document. Our empirical results also demonstrated that learning a multiple-field document representation is superior to aggregating retrieval scores from matching the query with different fields. We further showed that field-level masking and dropout are useful for handling fields with variable number of text instances and avoiding over-dependence on high-accuracy fields, respectively.

This work smooths the path towards pursuing several research directions in the future. For instance, many retrieval tasks based on semi-structured documents, such as academic search, XML retrieval, product search, expert finding, etc. can benefit from the \nrmf~framework for improving the retrieval performance. Another possible future direction is to extend the \nrmf~framework by considering lexical matching of query and document fields. We can also explore incorporating query independent features, such as, PageRank score, into our framework. Finally, our findings may be applicable to non-ranking tasks, including document classification, spam detection, and document filtering.

\paragraph{\textbf{Acknowledgements.}} \small This work was supported in part by the Center for Intelligent Information Retrieval. Any opinions, findings and conclusions or recommendations expressed in this material are those of the authors and do not necessarily reflect those of the sponsor.



\begin{thebibliography}{00}


\ifx \showCODEN    \undefined \def \showCODEN     #1{\unskip}     \fi
\ifx \showDOI      \undefined \def \showDOI       #1{{\tt DOI:}\penalty0{#1}\ }
  \fi
\ifx \showISBNx    \undefined \def \showISBNx     #1{\unskip}     \fi
\ifx \showISBNxiii \undefined \def \showISBNxiii  #1{\unskip}     \fi
\ifx \showISSN     \undefined \def \showISSN      #1{\unskip}     \fi
\ifx \showLCCN     \undefined \def \showLCCN      #1{\unskip}     \fi
\ifx \shownote     \undefined \def \shownote      #1{#1}          \fi
\ifx \showarticletitle \undefined \def \showarticletitle #1{#1}   \fi
\ifx \showURL      \undefined \def \showURL       #1{#1}          \fi
\providecommand\bibfield[2]{#2}
\providecommand\bibinfo[2]{#2}
\providecommand\natexlab[1]{#1}
\providecommand\showeprint[2][]{arXiv:#2}

\bibitem[\protect\citeauthoryear{Agichtein, Brill, and Dumais}{Agichtein
  et~al\mbox{.}}{2006}]%
        {Agichtein:2006}
\bibfield{author}{\bibinfo{person}{Eugene Agichtein}, \bibinfo{person}{Eric
  Brill}, {and} \bibinfo{person}{Susan Dumais}.}
  \bibinfo{year}{2006}\natexlab{}.
\newblock \showarticletitle{Improving Web Search Ranking by Incorporating User
  Behavior Information}. In \bibinfo{booktitle}{{\em Proceedings of the 29th
  Annual International ACM SIGIR Conference on Research and Development in
  Information Retrieval}} {\em (\bibinfo{series}{SIGIR '06})}.
  \bibinfo{publisher}{ACM}, \bibinfo{address}{Seattle, Washington, USA},
  \bibinfo{pages}{19--26}.
\newblock
\showISBNx{1-59593-369-7}


\bibitem[\protect\citeauthoryear{Borisov, Markov, de~Rijke, and
  Serdyukov}{Borisov et~al\mbox{.}}{2016}]%
        {Borisov:2016}
\bibfield{author}{\bibinfo{person}{Alexey Borisov}, \bibinfo{person}{Ilya
  Markov}, \bibinfo{person}{Maarten de Rijke}, {and} \bibinfo{person}{Pavel
  Serdyukov}.} \bibinfo{year}{2016}\natexlab{}.
\newblock \showarticletitle{A Neural Click Model for Web Search}. In
  \bibinfo{booktitle}{{\em Proceedings of the 25th International Conference on
  World Wide Web}} {\em (\bibinfo{series}{WWW '16})}.
  \bibinfo{publisher}{IW3C2}, \bibinfo{address}{Montreal, Quebec, Canada},
  \bibinfo{pages}{531--541}.
\newblock
\showISBNx{978-1-4503-4143-1}


\bibitem[\protect\citeauthoryear{Burges, Ragno, and Le}{Burges
  et~al\mbox{.}}{2006}]%
        {Burges:2007}
\bibfield{author}{\bibinfo{person}{Christopher~J. Burges},
  \bibinfo{person}{Robert Ragno}, {and} \bibinfo{person}{Quoc~V. Le}.}
  \bibinfo{year}{2006}\natexlab{}.
\newblock \showarticletitle{Learning to Rank with Nonsmooth Cost Functions}.
\newblock In \bibinfo{booktitle}{{\em Advances in Neural Information Processing
  Systems 19}}. \bibinfo{publisher}{MIT Press}, \bibinfo{address}{Vancouver,
  B.C., Canada}, \bibinfo{pages}{193--200}.
\newblock


\bibitem[\protect\citeauthoryear{Burges}{Burges}{2010}]%
        {Burges:2010}
\bibfield{author}{\bibinfo{person}{Christopher J.~C. Burges}.}
  \bibinfo{year}{2010}\natexlab{}.
\newblock \bibinfo{booktitle}{{\em From {RankNet} to {LambdaRank} to
  {LambdaMART}: An Overview}}.
\newblock \bibinfo{type}{{T}echnical {R}eport}. \bibinfo{institution}{Microsoft
  Research}.
\newblock


\bibitem[\protect\citeauthoryear{Dehghani, Zamani, Severyn, Kamps, and
  Croft}{Dehghani et~al\mbox{.}}{2017}]%
        {Dehghani:2017}
\bibfield{author}{\bibinfo{person}{Mostafa Dehghani}, \bibinfo{person}{Hamed
  Zamani}, \bibinfo{person}{Aliaksei Severyn}, \bibinfo{person}{Jaap Kamps},
  {and} \bibinfo{person}{W.~Bruce Croft}.} \bibinfo{year}{2017}\natexlab{}.
\newblock \showarticletitle{Neural Ranking Models with Weak Supervision}. In
  \bibinfo{booktitle}{{\em Proceedings of the 40th International ACM SIGIR
  Conference on Research and Development in Information Retrieval}} {\em
  (\bibinfo{series}{SIGIR '17})}. \bibinfo{publisher}{ACM},
  \bibinfo{address}{Shinjuku, Tokyo, Japan}, \bibinfo{pages}{65--74}.
\newblock
\showISBNx{978-1-4503-5022-8}


\bibitem[\protect\citeauthoryear{Fuhr, Lalmas, Malik, and Kazai}{Fuhr
  et~al\mbox{.}}{2006}]%
        {Fuhr:2006}
\bibfield{author}{\bibinfo{person}{Norbert Fuhr}, \bibinfo{person}{Mounia
  Lalmas}, \bibinfo{person}{Saadia Malik}, {and} \bibinfo{person}{Gabriella
  Kazai}.} \bibinfo{year}{2006}\natexlab{}.
\newblock \bibinfo{booktitle}{{\em Advances in XML Information Retrieval and
  Evaluation: 4th International Workshop of the Initiative for the Evaluation
  of XML Retrieval}}.
\newblock \bibinfo{publisher}{Springer-Verlag New York, Inc.},
  \bibinfo{address}{Secaucus, NJ, USA}.
\newblock


\bibitem[\protect\citeauthoryear{Guo, Fan, Ai, and Croft}{Guo
  et~al\mbox{.}}{2016}]%
        {Guo:2016}
\bibfield{author}{\bibinfo{person}{Jiafeng Guo}, \bibinfo{person}{Yixing Fan},
  \bibinfo{person}{Qingyao Ai}, {and} \bibinfo{person}{W.~Bruce Croft}.}
  \bibinfo{year}{2016}\natexlab{}.
\newblock \showarticletitle{A Deep Relevance Matching Model for Ad-hoc
  Retrieval}. In \bibinfo{booktitle}{{\em Proceedings of the 25th ACM
  International on Conference on Information and Knowledge Management}} {\em
  (\bibinfo{series}{CIKM '16})}. \bibinfo{publisher}{ACM},
  \bibinfo{address}{Indianapolis, Indiana, USA}, \bibinfo{pages}{55--64}.
\newblock
\showISBNx{978-1-4503-4073-1}


\bibitem[\protect\citeauthoryear{Huang, He, Gao, Deng, Acero, and Heck}{Huang
  et~al\mbox{.}}{2013}]%
        {Huang:2013}
\bibfield{author}{\bibinfo{person}{Po-Sen Huang}, \bibinfo{person}{Xiaodong
  He}, \bibinfo{person}{Jianfeng Gao}, \bibinfo{person}{Li Deng},
  \bibinfo{person}{Alex Acero}, {and} \bibinfo{person}{Larry Heck}.}
  \bibinfo{year}{2013}\natexlab{}.
\newblock \showarticletitle{Learning Deep Structured Semantic Models for Web
  Search Using Clickthrough Data}. In \bibinfo{booktitle}{{\em Proceedings of
  the 22nd ACM International Conference on Information \& Knowledge
  Management}} {\em (\bibinfo{series}{CIKM '13})}. \bibinfo{publisher}{ACM},
  \bibinfo{address}{San Francisco, California, USA},
  \bibinfo{pages}{2333--2338}.
\newblock
\showISBNx{978-1-4503-2263-8}


\bibitem[\protect\citeauthoryear{Kim, Xue, and Croft}{Kim
  et~al\mbox{.}}{2009}]%
        {Kim:2009}
\bibfield{author}{\bibinfo{person}{Jinyoung Kim}, \bibinfo{person}{Xiaobing
  Xue}, {and} \bibinfo{person}{W.~Bruce Croft}.}
  \bibinfo{year}{2009}\natexlab{}.
\newblock \showarticletitle{A Probabilistic Retrieval Model for Semistructured
  Data}. In \bibinfo{booktitle}{{\em Proceedings of the 31th European
  Conference on IR Research on Advances in Information Retrieval}} {\em
  (\bibinfo{series}{ECIR '09})}. \bibinfo{publisher}{Springer-Verlag},
  \bibinfo{address}{Toulouse, France}, \bibinfo{pages}{228--239}.
\newblock
\showISBNx{978-3-642-00957-0}


\bibitem[\protect\citeauthoryear{Kim and Croft}{Kim and Croft}{2012}]%
        {Kim:2012}
\bibfield{author}{\bibinfo{person}{Jin~Young Kim} {and}
  \bibinfo{person}{W.~Bruce Croft}.} \bibinfo{year}{2012}\natexlab{}.
\newblock \showarticletitle{A Field Relevance Model for Structured Document
  Retrieval}. In \bibinfo{booktitle}{{\em Proceedings of the 34th European
  Conference on Advances in Information Retrieval}} {\em
  (\bibinfo{series}{ECIR'12})}. \bibinfo{publisher}{Springer-Verlag},
  \bibinfo{address}{Barcelona, Spain}, \bibinfo{pages}{97--108}.
\newblock
\showISBNx{978-3-642-28996-5}


\bibitem[\protect\citeauthoryear{Kingma and Ba}{Kingma and Ba}{2014}]%
        {Kingma:2014}
\bibfield{author}{\bibinfo{person}{Diederik~P. Kingma} {and}
  \bibinfo{person}{Jimmy Ba}.} \bibinfo{year}{2014}\natexlab{}.
\newblock \showarticletitle{Adam: A Method for Stochastic Optimization}. In
  \bibinfo{booktitle}{{\em Proceedings of the third International Conference on
  Learning Representations}} {\em (\bibinfo{series}{ICLR '14})}.
  \bibinfo{address}{Banff, Canada}.
\newblock


\bibitem[\protect\citeauthoryear{Lalmas and Tombros}{Lalmas and
  Tombros}{2007}]%
        {Lalmas:2007}
\bibfield{author}{\bibinfo{person}{Mounia Lalmas} {and}
  \bibinfo{person}{Anastasios Tombros}.} \bibinfo{year}{2007}\natexlab{}.
\newblock \showarticletitle{Evaluating XML Retrieval Effectiveness at INEX}.
\newblock \bibinfo{journal}{{\em SIGIR Forum\/}} \bibinfo{volume}{41},
  \bibinfo{number}{1} (\bibinfo{date}{June} \bibinfo{year}{2007}),
  \bibinfo{pages}{40--57}.
\newblock
\showISSN{0163-5840}


\bibitem[\protect\citeauthoryear{Lu and Li}{Lu and Li}{2013}]%
        {Lu:2013}
\bibfield{author}{\bibinfo{person}{Zhengdong Lu} {and} \bibinfo{person}{Hang
  Li}.} \bibinfo{year}{2013}\natexlab{}.
\newblock \showarticletitle{A Deep Architecture for Matching Short Texts}. In
  \bibinfo{booktitle}{{\em Advances in Neural Information Processing Systems
  26}} {\em (\bibinfo{series}{NIPS '13})}. \bibinfo{address}{Lake Tahoe, CA,
  USA}, \bibinfo{pages}{1367--1375}.
\newblock


\bibitem[\protect\citeauthoryear{Mitra and Craswell}{Mitra and
  Craswell}{2018}]%
        {mitra2017introduction}
\bibfield{author}{\bibinfo{person}{Bhaskar Mitra} {and} \bibinfo{person}{Nick
  Craswell}.} \bibinfo{year}{2018}\natexlab{}.
\newblock \showarticletitle{An Introduction to Neural Information Retrieval}.
\newblock \bibinfo{journal}{{\em Foundations and Trends{\textregistered} in
  Information Retrieval (to appear)\/}} (\bibinfo{year}{2018}).
\newblock


\bibitem[\protect\citeauthoryear{Mitra, Diaz, and Craswell}{Mitra
  et~al\mbox{.}}{2017}]%
        {Mitra:2017}
\bibfield{author}{\bibinfo{person}{Bhaskar Mitra}, \bibinfo{person}{Fernando
  Diaz}, {and} \bibinfo{person}{Nick Craswell}.}
  \bibinfo{year}{2017}\natexlab{}.
\newblock \showarticletitle{Learning to Match Using Local and Distributed
  Representations of Text for Web Search}. In \bibinfo{booktitle}{{\em
  Proceedings of the 26th International Conference on World Wide Web}} {\em
  (\bibinfo{series}{WWW '17})}. \bibinfo{publisher}{IW3C2},
  \bibinfo{address}{Perth, Australia}, \bibinfo{pages}{1291--1299}.
\newblock
\showISBNx{978-1-4503-4913-0}


\bibitem[\protect\citeauthoryear{Myaeng, Jang, Kim, and Zhoo}{Myaeng
  et~al\mbox{.}}{1998}]%
        {Myaeng:1998}
\bibfield{author}{\bibinfo{person}{Sung~Hyon Myaeng}, \bibinfo{person}{Don-Hyun
  Jang}, \bibinfo{person}{Mun-Seok Kim}, {and} \bibinfo{person}{Zong-Cheol
  Zhoo}.} \bibinfo{year}{1998}\natexlab{}.
\newblock \showarticletitle{A Flexible Model for Retrieval of SGML Documents}.
  In \bibinfo{booktitle}{{\em Proceedings of the 21st Annual International ACM
  SIGIR Conference on Research and Development in Information Retrieval}} {\em
  (\bibinfo{series}{SIGIR '98})}. \bibinfo{publisher}{ACM},
  \bibinfo{address}{Melbourne, Australia}, \bibinfo{pages}{138--145}.
\newblock


\bibitem[\protect\citeauthoryear{Nalisnick, Mitra, Craswell, and
  Caruana}{Nalisnick et~al\mbox{.}}{2016}]%
        {nalisnick2016improving}
\bibfield{author}{\bibinfo{person}{Eric Nalisnick}, \bibinfo{person}{Bhaskar
  Mitra}, \bibinfo{person}{Nick Craswell}, {and} \bibinfo{person}{Rich
  Caruana}.} \bibinfo{year}{2016}\natexlab{}.
\newblock \showarticletitle{Improving Document Ranking with Dual Word
  Embeddings}. In \bibinfo{booktitle}{{\em Proceedings of the 25th
  International Conference Companion on World Wide Web}} {\em
  (\bibinfo{series}{WWW '16 Companion})}. \bibinfo{publisher}{IW3C2},
  \bibinfo{address}{Montreal, Quebec, Canada}, \bibinfo{pages}{83--84}.
\newblock


\bibitem[\protect\citeauthoryear{Ogilvie and Callan}{Ogilvie and
  Callan}{2003}]%
        {Ogilvie:2003}
\bibfield{author}{\bibinfo{person}{Paul Ogilvie} {and} \bibinfo{person}{Jamie
  Callan}.} \bibinfo{year}{2003}\natexlab{}.
\newblock \showarticletitle{Combining Document Representations for Known-item
  Search}. In \bibinfo{booktitle}{{\em Proceedings of the 26th Annual
  International ACM SIGIR Conference on Research and Development in Informaion
  Retrieval}} {\em (\bibinfo{series}{SIGIR '03})}. \bibinfo{publisher}{ACM},
  \bibinfo{address}{Toronto, Canada}, \bibinfo{pages}{143--150}.
\newblock


\bibitem[\protect\citeauthoryear{Piwowarski and Gallinari}{Piwowarski and
  Gallinari}{2003}]%
        {Piwowarski:2003}
\bibfield{author}{\bibinfo{person}{Benjamin Piwowarski} {and}
  \bibinfo{person}{Patrick Gallinari}.} \bibinfo{year}{2003}\natexlab{}.
\newblock \showarticletitle{A machine learning model for information retrieval
  with structured documents}.
\newblock \bibinfo{journal}{{\em Machine Learning and Data Mining in Pattern
  Recognition\/}} (\bibinfo{year}{2003}), \bibinfo{pages}{425--438}.
\newblock


\bibitem[\protect\citeauthoryear{Qin, Liu, Xu, and Li}{Qin
  et~al\mbox{.}}{2010}]%
        {Qin:2010}
\bibfield{author}{\bibinfo{person}{Tao Qin}, \bibinfo{person}{Tie-Yan Liu},
  \bibinfo{person}{Jun Xu}, {and} \bibinfo{person}{Hang Li}.}
  \bibinfo{year}{2010}\natexlab{}.
\newblock \showarticletitle{LETOR: A Benchmark Collection for Research on
  Learning to Rank for Information Retrieval}.
\newblock \bibinfo{journal}{{\em Inf. Retr.\/}} \bibinfo{volume}{13},
  \bibinfo{number}{4} (\bibinfo{date}{Aug.} \bibinfo{year}{2010}),
  \bibinfo{pages}{346--374}.
\newblock
\showISSN{1386-4564}


\bibitem[\protect\citeauthoryear{Robertson, Zaragoza, and Taylor}{Robertson
  et~al\mbox{.}}{2004}]%
        {Robertson:2004}
\bibfield{author}{\bibinfo{person}{Stephen Robertson}, \bibinfo{person}{Hugo
  Zaragoza}, {and} \bibinfo{person}{Michael Taylor}.}
  \bibinfo{year}{2004}\natexlab{}.
\newblock \showarticletitle{Simple BM25 Extension to Multiple Weighted Fields}.
  In \bibinfo{booktitle}{{\em Proceedings of the Thirteenth ACM International
  Conference on Information and Knowledge Management}} {\em
  (\bibinfo{series}{CIKM '04})}. \bibinfo{publisher}{ACM},
  \bibinfo{address}{Washington, D.C., USA}, \bibinfo{pages}{42--49}.
\newblock
\showISBNx{1-58113-874-1}


\bibitem[\protect\citeauthoryear{Robertson, Walker, Jones, Hancock-Beaulieu,
  and Gatford}{Robertson et~al\mbox{.}}{1994}]%
        {Robertson:1994}
\bibfield{author}{\bibinfo{person}{S.~E. Robertson}, \bibinfo{person}{E.
  Walker}, \bibinfo{person}{S. Jones}, \bibinfo{person}{M.~M.
  Hancock-Beaulieu}, {and} \bibinfo{person}{M. Gatford}.}
  \bibinfo{year}{1994}\natexlab{}.
\newblock \showarticletitle{Okapi at TREC-3}. In \bibinfo{booktitle}{{\em
  Proceedings of the third Text Retrieval Conference}} {\em
  (\bibinfo{series}{TREC '94})}.
\newblock


\bibitem[\protect\citeauthoryear{Shen, He, Gao, Deng, and Mesnil}{Shen
  et~al\mbox{.}}{2014}]%
        {Shen:2014}
\bibfield{author}{\bibinfo{person}{Yelong Shen}, \bibinfo{person}{Xiaodong He},
  \bibinfo{person}{Jianfeng Gao}, \bibinfo{person}{Li Deng}, {and}
  \bibinfo{person}{Gr{\'e}goire Mesnil}.} \bibinfo{year}{2014}\natexlab{}.
\newblock \showarticletitle{A Latent Semantic Model with Convolutional-Pooling
  Structure for Information Retrieval}. In \bibinfo{booktitle}{{\em Proceedings
  of the 23rd ACM International Conference on Conference on Information and
  Knowledge Management}} {\em (\bibinfo{series}{CIKM '14})}.
  \bibinfo{publisher}{ACM}, \bibinfo{address}{Shanghai, China},
  \bibinfo{pages}{101--110}.
\newblock
\showISBNx{978-1-4503-2598-1}


\bibitem[\protect\citeauthoryear{Srivastava, Hinton, Krizhevsky, Sutskever, and
  Salakhutdinov}{Srivastava et~al\mbox{.}}{2014}]%
        {Srivastava:2014}
\bibfield{author}{\bibinfo{person}{Nitish Srivastava},
  \bibinfo{person}{Geoffrey Hinton}, \bibinfo{person}{Alex Krizhevsky},
  \bibinfo{person}{Ilya Sutskever}, {and} \bibinfo{person}{Ruslan
  Salakhutdinov}.} \bibinfo{year}{2014}\natexlab{}.
\newblock \showarticletitle{Dropout: A Simple Way to Prevent Neural Networks
  from Overfitting}.
\newblock \bibinfo{journal}{{\em Journal of Machine Learning Research\/}}
  \bibinfo{volume}{15} (\bibinfo{year}{2014}), \bibinfo{pages}{1929--1958}.
\newblock


\bibitem[\protect\citeauthoryear{Svore and Burges}{Svore and Burges}{2009}]%
        {Svore:2009}
\bibfield{author}{\bibinfo{person}{Krysta~M. Svore} {and}
  \bibinfo{person}{Christopher~J.C. Burges}.} \bibinfo{year}{2009}\natexlab{}.
\newblock \showarticletitle{A Machine Learning Approach for Improved BM25
  Retrieval}. In \bibinfo{booktitle}{{\em Proceedings of the 18th ACM
  Conference on Information and Knowledge Management}} {\em
  (\bibinfo{series}{CIKM '09})}. \bibinfo{publisher}{ACM},
  \bibinfo{address}{Hong Kong, China}, \bibinfo{pages}{1811--1814}.
\newblock
\showISBNx{978-1-60558-512-3}


\bibitem[\protect\citeauthoryear{Wilkinson}{Wilkinson}{1994}]%
        {Wilkinson:1994}
\bibfield{author}{\bibinfo{person}{Ross Wilkinson}.}
  \bibinfo{year}{1994}\natexlab{}.
\newblock \showarticletitle{Effective Retrieval of Structured Documents}. In
  \bibinfo{booktitle}{{\em Proceedings of the 17th Annual International ACM
  SIGIR Conference on Research and Development in Information Retrieval}} {\em
  (\bibinfo{series}{SIGIR '94})}. \bibinfo{publisher}{Springer-Verlag New York,
  Inc.}, \bibinfo{address}{Dublin, Ireland}, \bibinfo{pages}{311--317}.
\newblock
\showISBNx{0-387-19889-X}


\bibitem[\protect\citeauthoryear{Xiong, Dai, Callan, Liu, and Power}{Xiong
  et~al\mbox{.}}{2017}]%
        {Xiong:2017}
\bibfield{author}{\bibinfo{person}{Chenyan Xiong}, \bibinfo{person}{Zhuyun
  Dai}, \bibinfo{person}{Jamie Callan}, \bibinfo{person}{Zhiyuan Liu}, {and}
  \bibinfo{person}{Russell Power}.} \bibinfo{year}{2017}\natexlab{}.
\newblock \showarticletitle{End-to-End Neural Ad-hoc Ranking with Kernel
  Pooling}. In \bibinfo{booktitle}{{\em Proceedings of the 40th International
  ACM SIGIR Conference on Research and Development in Information Retrieval}}
  {\em (\bibinfo{series}{SIGIR '17})}. \bibinfo{publisher}{ACM},
  \bibinfo{address}{Shinjuku, Tokyo, Japan}, \bibinfo{pages}{55--64}.
\newblock
\showISBNx{978-1-4503-5022-8}


\bibitem[\protect\citeauthoryear{Xue, Zeng, Chen, Yu, Ma, Xi, and Fan}{Xue
  et~al\mbox{.}}{2004}]%
        {Xue:2004}
\bibfield{author}{\bibinfo{person}{Gui-Rong Xue}, \bibinfo{person}{Hua-Jun
  Zeng}, \bibinfo{person}{Zheng Chen}, \bibinfo{person}{Yong Yu},
  \bibinfo{person}{Wei-Ying Ma}, \bibinfo{person}{WenSi Xi}, {and}
  \bibinfo{person}{WeiGuo Fan}.} \bibinfo{year}{2004}\natexlab{}.
\newblock \showarticletitle{Optimizing Web Search Using Web Click-through
  Data}. In \bibinfo{booktitle}{{\em Proceedings of the Thirteenth ACM
  International Conference on Information and Knowledge Management}} {\em
  (\bibinfo{series}{CIKM '04})}. \bibinfo{publisher}{ACM},
  \bibinfo{address}{Washington, D.C., USA}, \bibinfo{pages}{118--126}.
\newblock
\showISBNx{1-58113-874-1}


\bibitem[\protect\citeauthoryear{Yang, Ai, Guo, and Croft}{Yang
  et~al\mbox{.}}{2016}]%
        {Yang:2016}
\bibfield{author}{\bibinfo{person}{Liu Yang}, \bibinfo{person}{Qingyao Ai},
  \bibinfo{person}{Jiafeng Guo}, {and} \bibinfo{person}{W.~Bruce Croft}.}
  \bibinfo{year}{2016}\natexlab{}.
\newblock \showarticletitle{aNMM: Ranking Short Answer Texts with
  Attention-Based Neural Matching Model}. In \bibinfo{booktitle}{{\em
  Proceedings of the 25th ACM International on Conference on Information and
  Knowledge Management}} {\em (\bibinfo{series}{CIKM '16})}.
  \bibinfo{publisher}{ACM}, \bibinfo{address}{Indianapolis, Indiana, USA},
  \bibinfo{pages}{287--296}.
\newblock
\showISBNx{978-1-4503-4073-1}


\bibitem[\protect\citeauthoryear{Zamani, Bendersky, Wang, and Zhang}{Zamani
  et~al\mbox{.}}{2017}]%
        {Zamani:2017b}
\bibfield{author}{\bibinfo{person}{Hamed Zamani}, \bibinfo{person}{Michael
  Bendersky}, \bibinfo{person}{Xuanhui Wang}, {and} \bibinfo{person}{Mingyang
  Zhang}.} \bibinfo{year}{2017}\natexlab{}.
\newblock \showarticletitle{Situational Context for Ranking in Personal
  Search}. In \bibinfo{booktitle}{{\em Proceedings of the 26th International
  Conference on World Wide Web}} {\em (\bibinfo{series}{WWW '17})}.
  \bibinfo{publisher}{IW3C2}, \bibinfo{address}{Perth, Australia},
  \bibinfo{pages}{1531--1540}.
\newblock
\showISBNx{978-1-4503-4913-0}


\bibitem[\protect\citeauthoryear{Zamani and Croft}{Zamani and Croft}{2016}]%
        {Zamani:2016}
\bibfield{author}{\bibinfo{person}{Hamed Zamani} {and}
  \bibinfo{person}{W.~Bruce Croft}.} \bibinfo{year}{2016}\natexlab{}.
\newblock \showarticletitle{Estimating Embedding Vectors for Queries}. In
  \bibinfo{booktitle}{{\em Proceedings of the 2016 ACM International Conference
  on the Theory of Information Retrieval}} {\em (\bibinfo{series}{ICTIR '16})}.
  \bibinfo{publisher}{ACM}, \bibinfo{address}{Newark, Delaware, USA},
  \bibinfo{pages}{123--132}.
\newblock


\bibitem[\protect\citeauthoryear{Zamani and Croft}{Zamani and Croft}{2017}]%
        {Zamani:2017a}
\bibfield{author}{\bibinfo{person}{Hamed Zamani} {and}
  \bibinfo{person}{W.~Bruce Croft}.} \bibinfo{year}{2017}\natexlab{}.
\newblock \showarticletitle{Relevance-based Word Embedding}. In
  \bibinfo{booktitle}{{\em Proceedings of the 40th International ACM SIGIR
  Conference on Research and Development in Information Retrieval}} {\em
  (\bibinfo{series}{SIGIR '17})}. \bibinfo{publisher}{ACM},
  \bibinfo{address}{Shinjuku, Tokyo, Japan}, \bibinfo{pages}{505--514}.
\newblock


\end{thebibliography}
\end{document}